\documentclass[11pt,twoside]{osajnl}

\journal{josaa} 
\usepackage{wrapfig}
\usepackage{subcaption}

\setboolean{shortarticle}{false}

\title{Measurements of Optical Scatter Versus Annealing Temperature for Amorphous Ta$_2$O$_5$ and TiO$_2$:Ta$_2$O$_5$ Thin Films}

\author[1,2,a]{Elenna M. Capote}
\author[1]{Amy Gleckl}
\author[1]{Jazlyn Guerrero}
\author[1]{Michael Rezac}
\author[1]{Robert Wright}
\author[1,b]{Joshua R. Smith}

\affil[1]{Gravitational-Wave Physics and Astronomy Center, California State University, Fullerton, Fullerton, CA, USA}

\affil[2]{Department of Physics, Syracuse University, Syracuse, NY, USA}

\affil[a]{ecapote@syr.edu}
\affil[b]{josmith@fullerton.edu}

\begin{abstract}
Optical coatings formed from amorphous oxide thin films have many applications in precision measurements. The Advanced Laser Interferometer Gravitational-Wave Observatory (LIGO) and Advanced Virgo use coatings of SiO$_2$ (silica) and TiO$_2$:Ta$_2$O$_5$ (titania doped tantala) and post-deposition annealing to 500$^\circ$C to achieve low thermal noise and low optical absorption. Optical scattering by these coatings is a key limit to the detectors' sensitivity. This paper describes optical scattering measurements for single-layer ion-beam-sputtered thin films on fused silica substrates: two samples of Ta$_2$O$_5$ and two of TiO$_2$:Ta$_2$O$_5$. Using an imaging scatterometer at a fixed scattering angle of 12.8$^\circ$, in-situ changes in the optical scatter of each sample were assessed during post-deposition annealing to 500$^\circ$C in vacuum. The scatter of three of the four coated optics was observed to decrease during the annealing process, by 25-30$\%$ for tantala and up to 74$\%$ for titania-doped tantala, while scatter from the fourth sample held constant. Angle-resolved scatter measurements performed before and after vacuum annealing suggests some improvement in three of four samples. These results demonstrate that post-deposition high-temperature annealing of single-layer tantala and titania-doped tantala thin films in vacuum does not lead to an increase in scatter, and may actually improve their scatter.

\end{abstract}

\setboolean{displaycopyright}{true}

\begin{document}

\maketitle

\section{Introduction}

Amorphous oxide multilayer thin films are important for a number of applications such as atomic clocks and laser interferometers~\cite{Harry:2011book}. Recent gravitational-wave discoveries~\cite{gw150914,gw170817,LIGOScientific:2018mvr} by the Advanced Laser Interferometer Gravitational Wave Observatory (LIGO)~\cite{TheLIGOScientific:2014jea} and Advanced Virgo~\cite{TheVirgo:2014hva} were enabled by low thermal noise and low optical absorption coatings on its core optics~\cite{coreoptics, PhysRevD.98.122001, Harry:2011book, Granata_2020}. These coatings are formed from alternating layers of low index of refraction SiO$_2$ and high index of refraction TiO$_2$:Ta$_2$O$_5$ (tantala, doped with about 20\% titania). 
To reduce the mechanical loss (which determines Brownian thermal noise) and the optical absorption of the TiO$_2$:Ta$_2$O$_5$ layers, the coatings are annealed post-deposition to 500$^\circ$C for 10 hours. As shown in Amato et. al. and Granata et. al., post-deposition annealing is an effective way to "erase deposition history" and ensure the coating is configured in a way to minimize mechanical loss for both Ta$_2$O$_5$ and TiO$_2$:Ta$_2$O$_5$~\cite{Amato_2018,Granata_2020}. The addition of the TiO$_2$ dopant also improves absorption and mechanical losses in the coating, while potentially inhibiting crystallization at lower temperatures. The in-air crystallization temperature of tantala has been shown to be 650$^{\circ}$C, and titania-doped tantala 700$^{\circ}$C~\cite{Fazio:2020yqa}. As shown in Fazio et. al., annealing TiO$_2$:Ta$_2$O$_5$ to 600$^\circ$C is ideal for decreasing mechanical loss and lowering optical absorption~\cite{Fazio:2020yqa}. 

However, the effects of annealing on the optical scattering of these coatings is less well understood. 
Noise from scattering is a limit to LIGO and Virgo, causing direct reduction of power in the Fabry-Perot arm cavities, scatter from primary mirrors hitting walls and baffles and re-entering~\cite{Accadia_2010, Flanagan_1994}, and direct reduction of the quality of quantum squeezed states~\cite{Kwee:2014vba}.
It is therefore desirable to improve upon the scatter performance of LIGO-Virgo optics for future detector upgrades. 
Advanced LIGO core optics exhibit an average scatter of 9.5 ppm, while Rayleigh-Rice perturbation theory predicts a total scatter of 2.4 ppm as a result of their surface roughness~\cite{coreoptics, Smith:19}. This excess scatter is due to point scattering of unknown origin. There is concern that the annealing of these optics could increase their optical scatter due to the formation and growth of crystals below the crystallization temperature, or other material changes~\cite{Glover:2018huy, Glover_2018}. This concern applies particularly to the high-index material of the multilayer stacks, TiO$_2$:Ta$_2$O$_5$, as it has the lower crystallization temperature.


Characterization of optical scatter from thin films has a rich tradition in optics (see, e.g., Refs.~\cite{Stover:2012book, Lequime2008, schroder2011angle, Harry:2011book}). Previous studies of optics for gravitational-wave detectors assessed the scatter only before and after the annealing process~\cite{Amra:2004cqa, coreoptics}. 
This paper presents, for single-layer Ta$_2$O$_5$ and TiO$_2$:Ta$_2$O$_5$ coatings, angle-resolved scatter measurements performed before and after annealing and, for the first time, in-situ results of optical scattering while annealing in vacuum. 
Such in-situ measurements provide the opportunity to better observe the time and manner in which morphological changes to the coating affect scatter. While this method is used for the purpose of studying coatings for gravitational-wave detectors, they can be applied to optical coatings with other applications.

\section{Quantifying Optical Scatter}{\label{sec:analysis}}

In this paper, two quantities will be used to characterize the scatter from coated optics: the Bidirectional Reflectance Distribution Function (BRDF) and the Integrated Scatter (IS). BRDF quantifies the amount of scattered light measured at a discrete solid angle, while IS is a measurement of the scatter over a large part of the entire hemisphere of backscattering.

BRDF is defined in Stover as

\begin{equation}
    BRDF = \frac{dP_s/d\Omega_s }{P_i \cos \theta_s} \cong \frac{P_s/\Omega_s }{P_i \cos \theta_s}
\end{equation}

where P$_i$ is the incident laser power and P$_s$ is the scattered light power measured at polar angle $\theta_s$ by the imaging system, which subtends a solid angle $\Omega_s$, from the normal to the optical surface, and in the plane of the laser beam \cite{Stover:2012book, Magana-Sandoval:12}.

Integrated Scatter is defined as the ratio of scattered power to incident power. IS is calculated by integrating a measurement of BRDF, multiplied by $\cos\theta_s$ over a full solid angle of scattering. The scattering over azimuthal angles is assumed to be isotropic \cite{Stover:2012book, Magana-Sandoval:12}. IS is therefore defined as

\begin{equation}
    IS = \frac{P_{s}}{P_{i}} = \int_0^{2\pi} \int_{\theta_{min}}^{\theta_{max}} BRDF(\theta_s) \cos \theta_s d\Omega_s
\end{equation}

Note that this differs from the Total Integrated Scatter or TIS, defined by Stover as the ratio of scattered power to specularly reflected power~\cite{Stover:2012book}, by a missing factor of R, the reflectivity of the optic. The integrated scatter becomes the total integrated scatter by dividing by R.

The angle-resolved scatterometer (ARS, defined in Section~\ref{sec:ars}) measures scatter at discrete angles between 3$^\circ$ < $\theta_s$ < 80$^\circ$, so the solid angle integral is approximated as a summation of the individual rings centered on the polar angles of measured scatter $\theta_s$~\cite{Magana-Sandoval:12}

\begin{equation}
    IS = 2 \pi (\cos \theta_1 - \cos \theta_2)BRDF(\theta_s)\cos \theta_s
\end{equation}

In order to calculate IS, measurements must be made over multiple angles, therefore the fixed scattering angle measurements in this paper made by the Vacuum Annealing Scatterometer (VAS, described in Section~\ref{sec:trs}) only produce a BRDF value in the analysis. The range of angles measured by the ARS varies slightly from measurement to measurement, depending on the angle of incidence that the sample is placed at, typically between 1.5$^\circ$ and 2.5$^\circ$.

\subsection{Multiple Region of Interest Analysis}

Both experiments described in this paper take images of scatter using a charged-coupled device, or CCD. For each CCD image taken, an elliptically-shaped region of interest (ROI) is defined around the area of significant measured scattered light. The counts of all pixels within the entire ROI are summed up and normalized by the exposure time and incident power. This value is multiplied by a calibration factor~\cite{Magana-Sandoval:12} previously determined (by comparison between a power meter and the CCD of the scatter from a diffuse reference) to give a measurement of BRDF.

However, each ROI selected will have some diffuse light or camera noise in addition to the scatter from the beamspot. To better estimate the optical scatter enclosed in the ROI, 6 concentric ROIs are defined in increasing size, as seen in Figure~\ref{fig:170811a_sidebyside}. The counts within each ROI are calculated and normalized. These 6 values are fit to a linear function versus pixel area which defines a y-intercept value that estimates the true enclosed BRDF without any extra diffuse light. This y-intercept value is multiplied with the calibration factor to define the BRDF. This type of analysis is performed on both the angle-resolved scatterometer and the vacuum annealing scatterometer data.

Multi-ROI analysis provides a more accurate measurement of scatter although it does have limitations. Certain samples have high scatter from their back surfaces that contaminates the the outer ROI at large angles. In this case, the linear fit may result in a too low or negative BRDF value. For this reason, some high scattering angles are rejected in the analysis.

\section{Samples}
\label{sec:samples}

\begin{figure}[h]
\begin{center}
    \includegraphics[height=10.1cm]{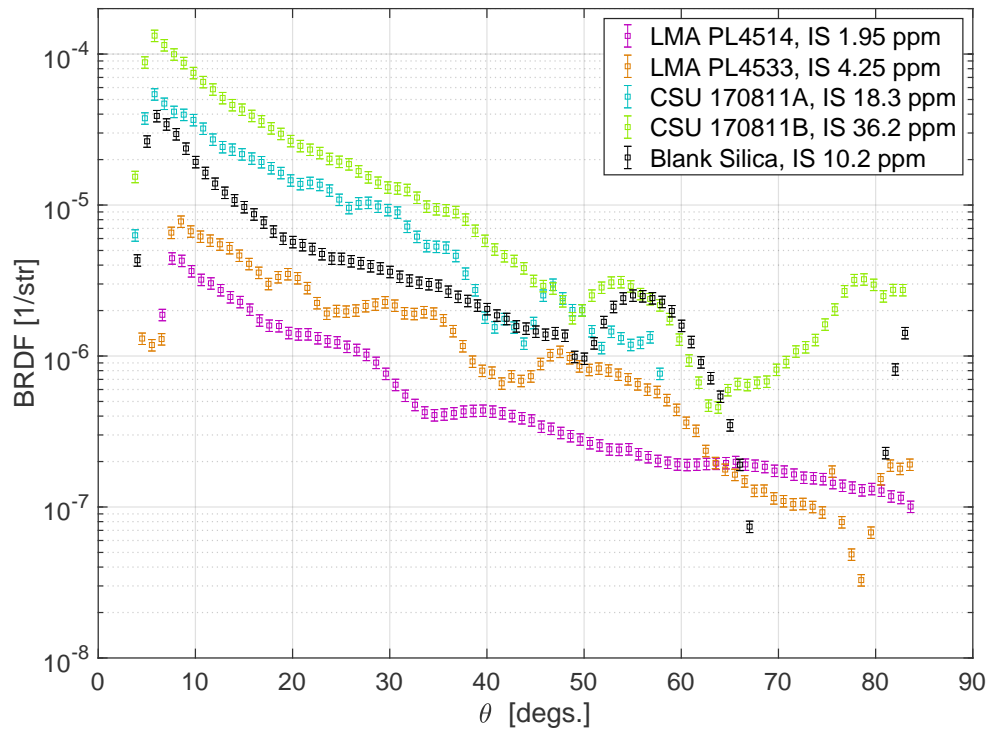}
    \caption{Optical scatter (BRDF) versus scattering angle obtained with the ARS for each of the four optics before they were annealed. Both LMA TiO$_2$:Ta$_2$O$_5$-coated samples show less scatter than the CSU Ta$_2$O$_5$-coated samples. An additional uncoated fused silica substrate (blank silica) is included for comparison.}
    \label{fig:pre-anneal}
\end{center}
\end{figure}

\begin{center}
\begin{table}
\centering
\begin{tabular}{c|c|c|c}
\hline
     Sample & Source & Coating & Thickness  \\
     \hline \hline
     PL4533 & LMA & Ti:Ta$_2$O$_5$ & 127 nm \\
     PL4514 & LMA & Ti:Ta$_2$O$_5$ & 127 nm \\
     170811a & CSU & Ta$_2$O$_5$ & 228 nm \\
     170811b & CSU & Ta$_2$O$_5$ & 226 nm \\
\end{tabular}
\caption{Information about the coated samples. The LMA samples were all coated in one group, so are uniform in sample thickness: 1064\,nm/(4n), where n$=2.09\pm0.01$~\cite{Granata_2020}.} 
\label{tab:sample-info}
\end{table}
\end{center}

The samples measured in this study are listed in Table~\ref{tab:sample-info}. The substrates are all 1-inch diameter fused silica substrates (Corning 7980) with superpolished front and back surfaces (by Coastline Optics) to a surface roughness of <1 {\AA} RMS (via Zygo 5500 optical profiler) and with flatness $<$2 waves peak-to-valley and $\leq 10^{-5}$ scratch dig surface quality, both in the central (80\%) area. The substrate barrels are also polished to reduce their scattering. These substrates were coated with single layers and not annealed. Two samples were coated with Ta$_2$O$_5$ at Colorado State University (CSU) and two with TiO$_2$:Ta$_2$O$_5$ at Laboratoire des Mat\'eriaux Avanc\'es (LMA). 


The optical scattering of the tantala-coated optics was measured in the angle-resolved scatterometer (described in Section~\ref{sec:ars}). The LMA samples were measured straight out of the package, but the CSU samples were first cleaned using a commercial polymer cleaning solution, First Contact, on the front and back sides to remove dust. The measured scatter of these samples is presented in Figure~\ref{fig:pre-anneal} and Table~\ref{tab:post-sample-info} as the BRDF and the IS. The BRDF is reported at the measured scattering angle closest to 12.8$^\circ$, as this is the scattering angle observed during annealing. These values are used to compare measurements before and after annealing. For sample 170811a, data at high angles (above $\theta_s$=60$^\circ$) has been removed because strong scatter from the back surface of the optic pollutes the region of interest used to estimate the front-surface scatter (see Section~\ref{sec:analysis}).

The BRDF and IS of the as-deposited TiO$_2$:Ta$_2$O$_5$ LMA samples are up to ten times less than that of the CSU Ta$_2$O$_5$ samples. 
Both CSU samples demonstrate higher scatter than the the uncoated fused silica substrate, while the LMA samples show lower scatter.  
Improved scatter after coating is typical as optical coatings can smooth over surface roughness of the substrates~(e.g., \cite{vander-hyde2015}).
These results are not altogether surprising, as CSU and LMA use different deposition methods, which can result in a different refractive index and absorption loss between two nominally similar samples~\cite{Fazio:2020yqa,Granata_2020}, and likely different scatter. Additionally, only the LMA coatings had titania dopant, which has been shown to improve absorption and mechanical losses~\cite{Fazio:2020yqa}, and may affect scatter. 



\section{Setup}

\subsection{Vacuum Annealing Scatterometer}\label{sec:trs}
\begin{figure}[h]
\begin{minipage}{0.5\textwidth}
	\includegraphics[width=1.0\textwidth]{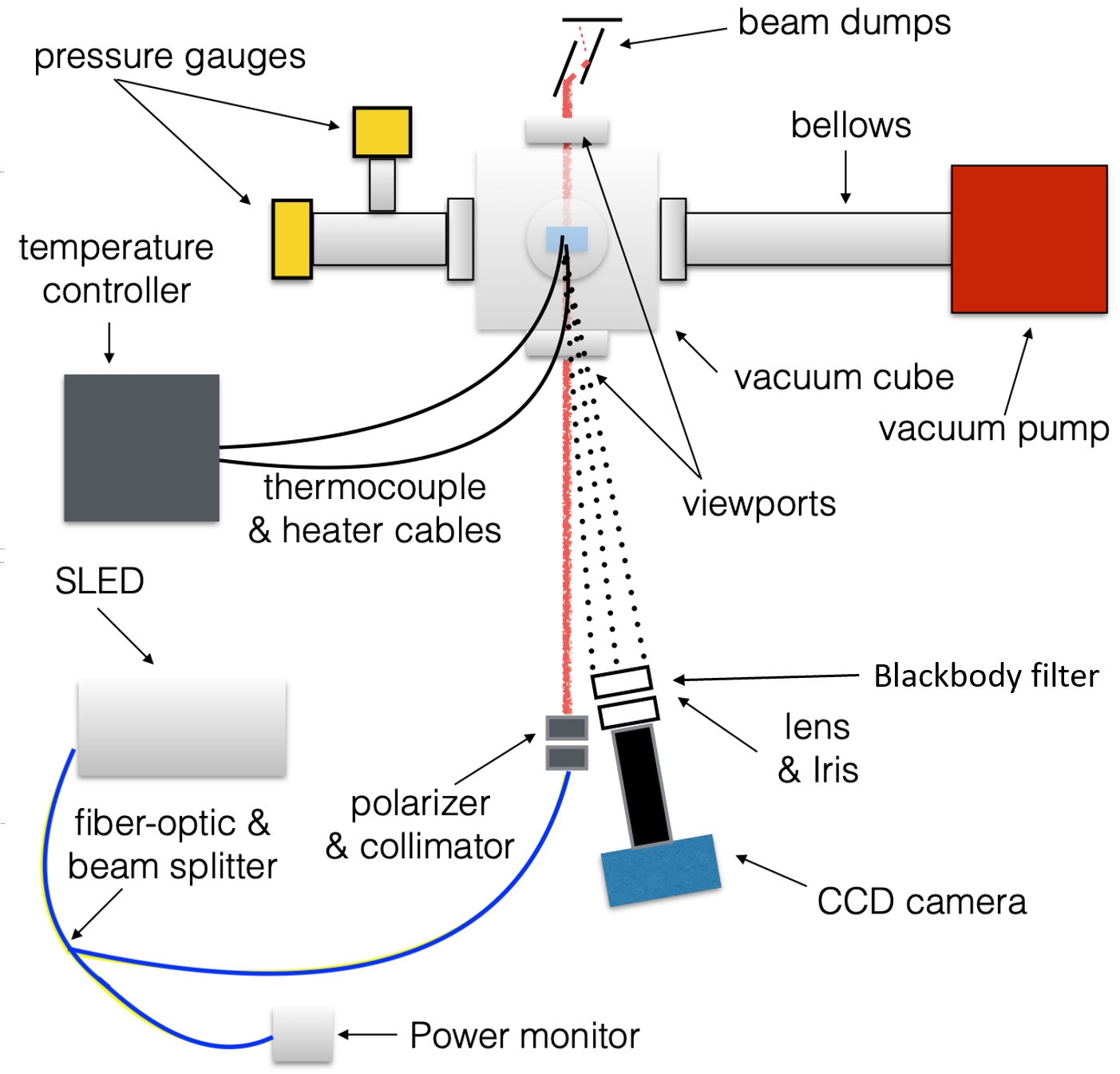}
    \caption{An overhead diagram of the Vacuum Annealing Scatterometer, including vacuum chamber, SLED, and imaging optics.}
    \label{fig:exp-layout}
\end{minipage}
\begin{minipage}{0.4\textwidth}
	\includegraphics[width=1.0\textwidth]{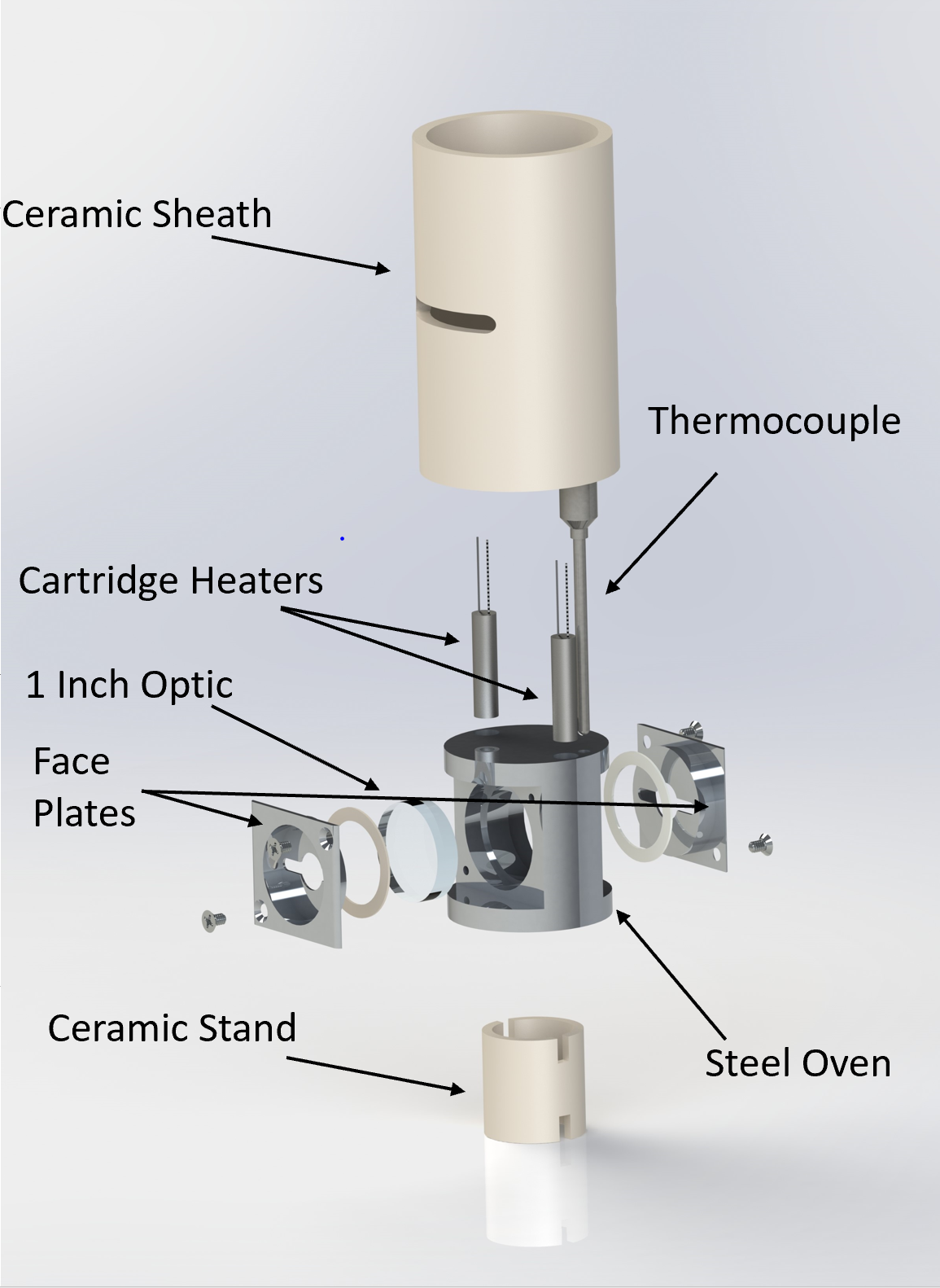}
    \caption{An "exploded" view of the stainless steel oven used to house the optic alongside the heaters and thermocouple. This is placed inside the vacuum cube shown in the experimental layout.}
    \label{fig:exploded-oven}
\end{minipage}
\end{figure}

The experimental apparatus used to measure optical scatter during the annealing process is shown in Figs.~\ref{fig:exp-layout} and ~\ref{fig:exploded-oven} and also described in Ref.~\cite{Smith:19}. An optic is placed in a vacuum chamber and held in place with ceramic rings on its front and back surface within a stainless steel oven, with stainless steel faceplates with a narrow slot to allow viewing of a small portion of the optical surface while reducing the area radiating to the colder surroundings. The oven is isolated from the vacuum chamber by a ceramic stand and ceramic sheath. Two 100 W Thermal Corporation CPN97393 stainless steel cartridge heaters and a type K Omega thermocouple are embedded on either side of the optic. 

The vacuum chamber is pumped down to below $1 \times10^{-5}$ torr before annealing begins. The sample is illuminated by a 1037 nm wavelength Super Luminescent Diode or SLED (Thorlabs S5FC1050P) through a fused silica anti-reflection coated vacuum viewport at normal incidence and with a beam diameter of 2.78\,mm. The incident power from the SLED ranges from 3 to 4 mW and the light is elliptically polarized. The heaters and thermocouple are wired to a programmable temperature control device (Watlow PM6R1CA). Typical temperature profiles follow a ramp rate of 1$^{\circ}$C/min, soak at the desired temperature for a 24 hours, and a ramp back down to room temperature at -1$^{\circ}$C/min.

Throughout this heating profile, a 4096$\times$4096 pixel CCD camera (Andor Apogee Alta F16) takes two images every 5 minutes, one "bright image" with the SLED illumination on and one "dark image" of the same exposure time but with the SLED turned off. The dark images are subtracted from the bright images to reduce blackbody radiation and ambient light in the images. For the temperatures achieved here, blackbody radiation completely dominates scattered light in the CCD's measurement band.  To account for this, a bandpass filter (Edmunds 1050nm/50nm) is placed at the entrance to the camera's tube to reduce blackbody radiation and room light at wavelengths far from 1037 nm. The images are formed on the CCD with a single lens and iris and taken at a fixed scattering angle of 12.8$^{\circ}$. The temperature profile, image capture, SLED switching, and data recording are all automated by a LabView Virtual Instrument. The entire experiment is placed on a seismically isolated optical bench inside a softwall cleanroom.

This experiment is designed to follow a similar protocol to the one used in the annealing of aLIGO coatings by LMA. While that exact procedure is proprietary, it is believed to be roughly comparable to ramp/soak parameters presented in other studies~\cite{Granata_2020, Fazio:2020yqa}. A distinction in this experiment from that protocol is that the annealing is performed in vacuum. 


\subsubsection{Incoherent Source}

This experiment makes use of an incoherent light source, a 1037 nm SLED, instead of a 1064 nm laser as used in LIGO, Virgo, and the ARS. The short coherence length of this SLED (8 microns) results in much less "twinkling" in the observed scatter than that from a laser. The incoherent source was chosen so that a more stable measurement of BRDF can be made, and small variations in the change in scatter can be more easily seen. The BRDF observed using incoherent light will differ from that using a laser by a small factor in general.

\subsubsection{VAS Capabilities}

The VAS is designed to measure the optical scatter of thin films in vacuum as they are annealed close to their crystallization temperature. Currently, the highest soak temperature reached with the cartridge heaters in vacuum is 520$^{\circ}$C. It is believed that the system can safely reach 560$^{\circ}$C under its current design, however, this has not yet been proven due lab closures associated with the pandemic. Within its current annealing temperature range, the VAS has successfully observed in-situ scatter increase, for samples with different substrates and coatings, due to either crystallization or delamination.


\subsection{Angle-Resolved Scatterometer}\label{sec:ars}

\begin{wrapfigure}{r}{7cm}
\includegraphics[scale=.5]{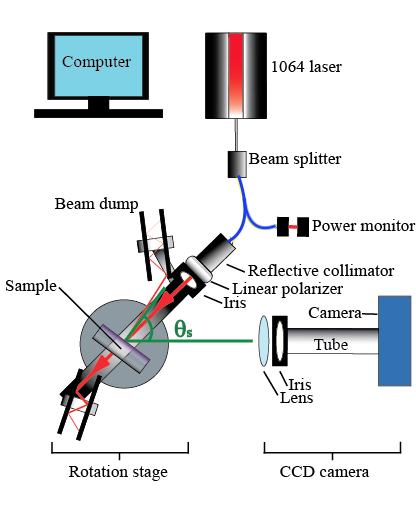}
\caption{An overhead diagram of the experimental set up for the angle-resolved scatterometer, including rotation stage, laser system, and optics.}
\label{fig:ars_setup}
\end{wrapfigure}

Angle-resolved scattering is an established technique for quantifying the scatter of coated optics~\cite{schroder2011angle, Lequime2008, Stover:2012book}.
Figure \ref{fig:ars_setup} shows the set up of the angle-resolved scatter experiment used to characterize scattering from optics as a function of scattering angle, $\theta_s$ (see Section~\ref{sec:analysis}). The system consists of an optical mount centered on a motorized rotation stage (Thorlabs HS Nano Rotator), a 1064nm laser (Innolight Mephisto S) incident on the sample with a fixed angle of incidence, and a CCD camera (Andor Apogee Alta F16), all mounted to a seismically isolated optical bench inside a softwall cleanroom. The laser outputs a beam to a 90:10 beamsplitter fiber optic cable, with 10\% going to a power monitoring photodiode. The 90\% cable is attached to a reflective collimator (Thorlabs RC08FC), fed through a linear polarizer to achieve horizontally polarized light, and further narrowed using an iris to approximately 6mm in diameter. Final incident power striking the optic's surface is typically 15mW. Initially, the optic, laser, and CCD camera are all in line and in the same plane. The optic and laser, which are connected by rails, are rotated in unison by the motorized rotation stage, controlled by LabView. The rotation stage rotates from 0$^{\circ}$ to 80$^{\circ}$ in increments of 1$^{\circ}$. At each angle, an image of scattered light from the optical surface is taken by the CCD camera, which remains fixed. The imaging system consists of a converging lens, an adjustable iris, and a CCD camera. An aluminum tube with an optical high-pass filter at its entrance is mounted to the face of the CCD camera to attenuate any room or stray laser light. Scatter is not observed until a minimum angle of 4$^{\circ}$, because prior to this angle the components of the laser (collimator, polarizing filter, and iris) block the view from the CCD camera. After a set of "bright images" have been taken from 0$^{\circ}$ to 80$^{\circ}$, the laser is turned off, and the process is repeated to take "dark images" of the same exposure length. During analysis, the dark images are subtracted from the bright images to ensure that room light is subtracted out.

\section{Results}

\begin{figure}[htbp]
\centering
\begin{minipage}{1.0\textwidth}
    \includegraphics[height=10.1cm]{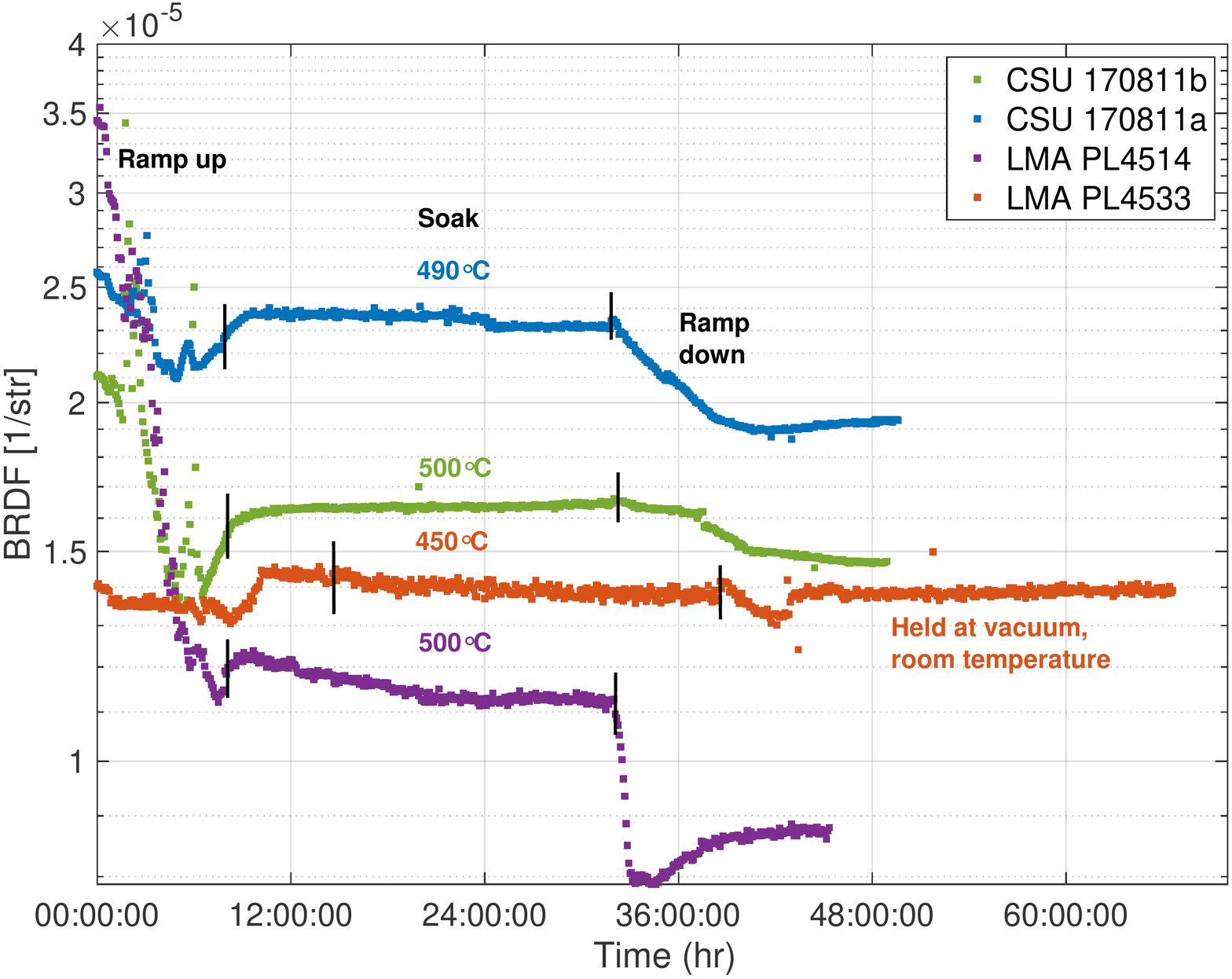}
    \end{minipage}
    \begin{minipage}{1.0\textwidth}
        \includegraphics[height=10.1cm]{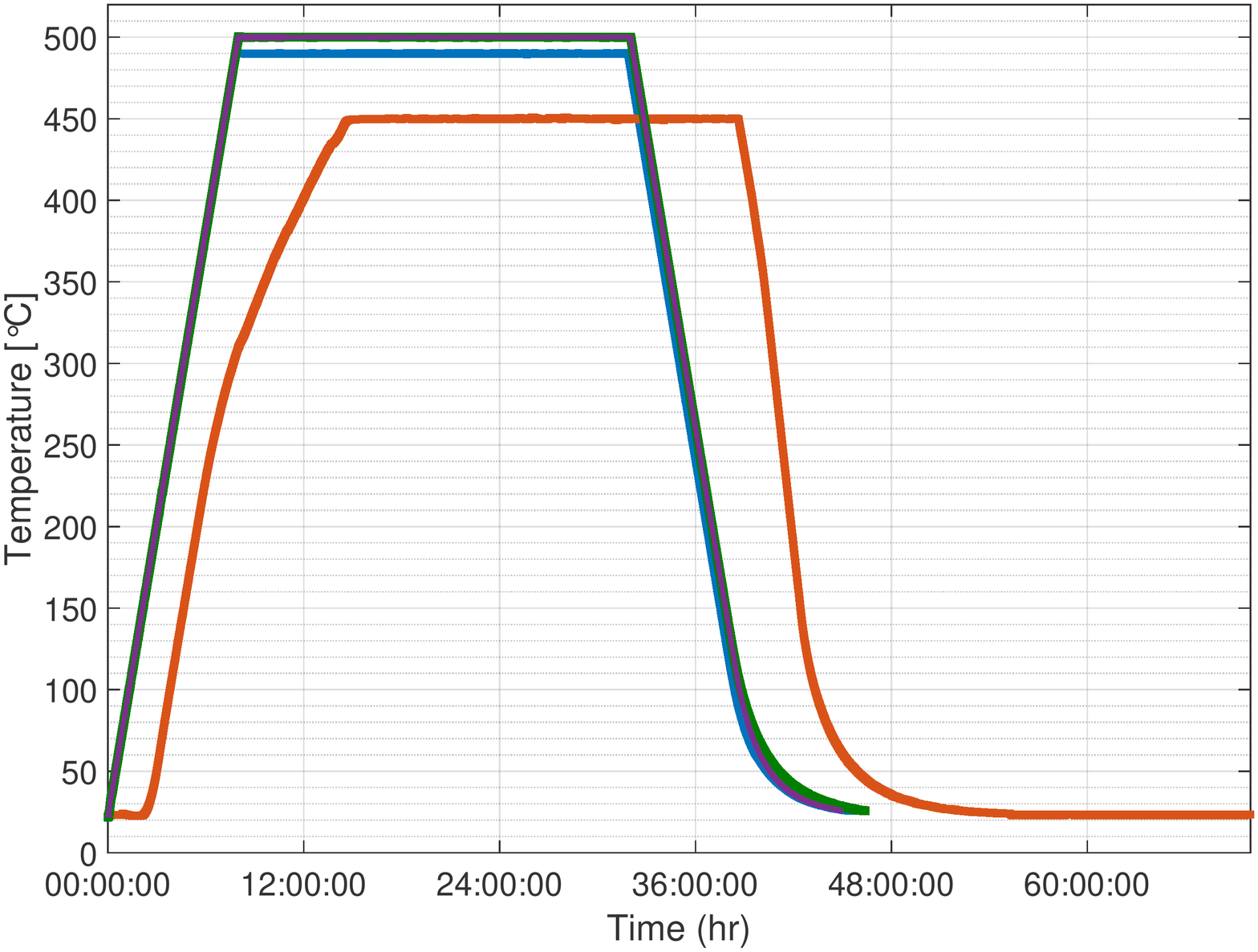}
    \end{minipage}
   
    \caption{Above: Log-y plot of BRDF versus time measured by the VAS for each sample, showing changes in measured scatter as each sample is annealed. Below: Corresponding measured temperatures, showing how each sample was ramped up to its annealing temperature, held for 24 hours (soak), and ramped down to room temperature. The black bars on the BRDF plot indicate when the 24-hour soak begins and ends and the soak temperature is labelled. For most samples, as the temperature increases, the scatter from each sample decreases a small amount. Scatter is more constant during the soak phase. Then scatter decreases again during the ramp back down to room temperature. One sample, PL4533, exhibited more constant scatter. This sample was held at room temperature in vacuum for an additional day.}
    \label{fig:annealing_brdf}
\end{figure}

\begin{figure}[htbp]
    \centering
    \includegraphics[width=1.0\textwidth]{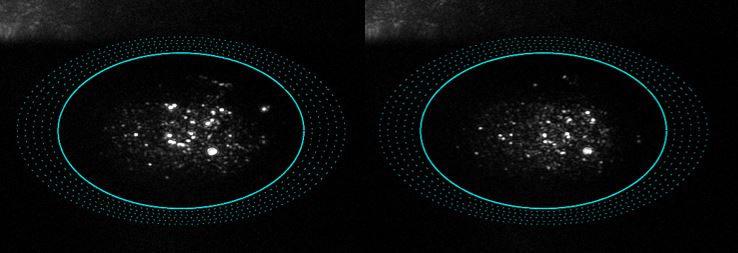}
    \caption{A side-by-side image of CSU sample 170811a while at room temperature at the start and end of annealing to 490$^{\circ}$C, measured at a scattering angle of 12.8$^{\circ}$ with the VAS. This comparison shows that certain scatter points have disappeared during the annealing process. The ovals around the scatter points represent the 6 regions of interest used to calculate BRDF, as described in Section~\ref{sec:analysis}. The innermost ROI defines an elliptical region that is 4.8\,mm$\times$4.0\,mm. Each subsequent ROI is 20\% larger in diameter than the previous.}
    \label{fig:170811a_sidebyside}
\end{figure}

Figure~\ref{fig:annealing_brdf} shows the scatter trend over time for $\theta_s = 12.8^{\circ}$ as the optics are annealed to temperatures ranging from 450$^{\circ}$C to 500$^{\circ}$C. The results demonstrate that to 500$^{\circ}$C, annealing in vacuum does not cause crystallization or other changes that increase the scatter from these coated optics, supporting the results from Fazio et. al~\cite{Fazio:2020yqa}. This holds true for both tantala (Ta$_2$O$_5$) and titania-doped tantala (TiO$_2$:Ta$_2$O$_5$) samples. In fact, the trends show that there is often a small decrease in the scatter during the ramp to the annealing temperature, and further decrease during the ramp down to room temperature. Images of the optics, such as the one shown in Figure~\ref{fig:170811a_sidebyside}, show that small scatter points disappear during the ramp, perhaps indicating that water, dust, or other contaminants are evaporated off during the heating.

Some increase in BRDF is seen as heater power increases towards the soak temperature. The cause of this is not fully understood. One possibility is that slight shifts and rotations in the set up cause the beamspot to shift over the optic. The wires connecting to the cartridge heaters likely apply torques to the optic that change with temperature. For later samples, a metal bar was installed at the bottom of the vacuum chamber that locked into grooves in the ceramic stand to prevent twisting. It was present for the annealing of sample PL4514, but not others.

These four samples were annealed without the ability to monitor the SLED power automatically throughout the experiment. Measurements of the SLED power were made by hand at the start, middle, and end of the experiment and indicated a fluctuation of SLED power that is at most 2$\%$. After sample PL4514 was annealed, a power monitor was installed, and used for other samples annealed at high temperatures. This power monitor confirmed that SLED power fluctuations are at most 1$\%$ for annealing up to 520$^\circ$C. Power monitoring also confirms there is no correlation between heater power and SLED power during annealing. The CCD camera temperature is also monitored throughout the experiment and stays cooled to around -18$^\circ$C during imaging.

\begin{figure}[htbp]
\begin{minipage}{0.5\textwidth}
   \includegraphics[width=1.0\textwidth]{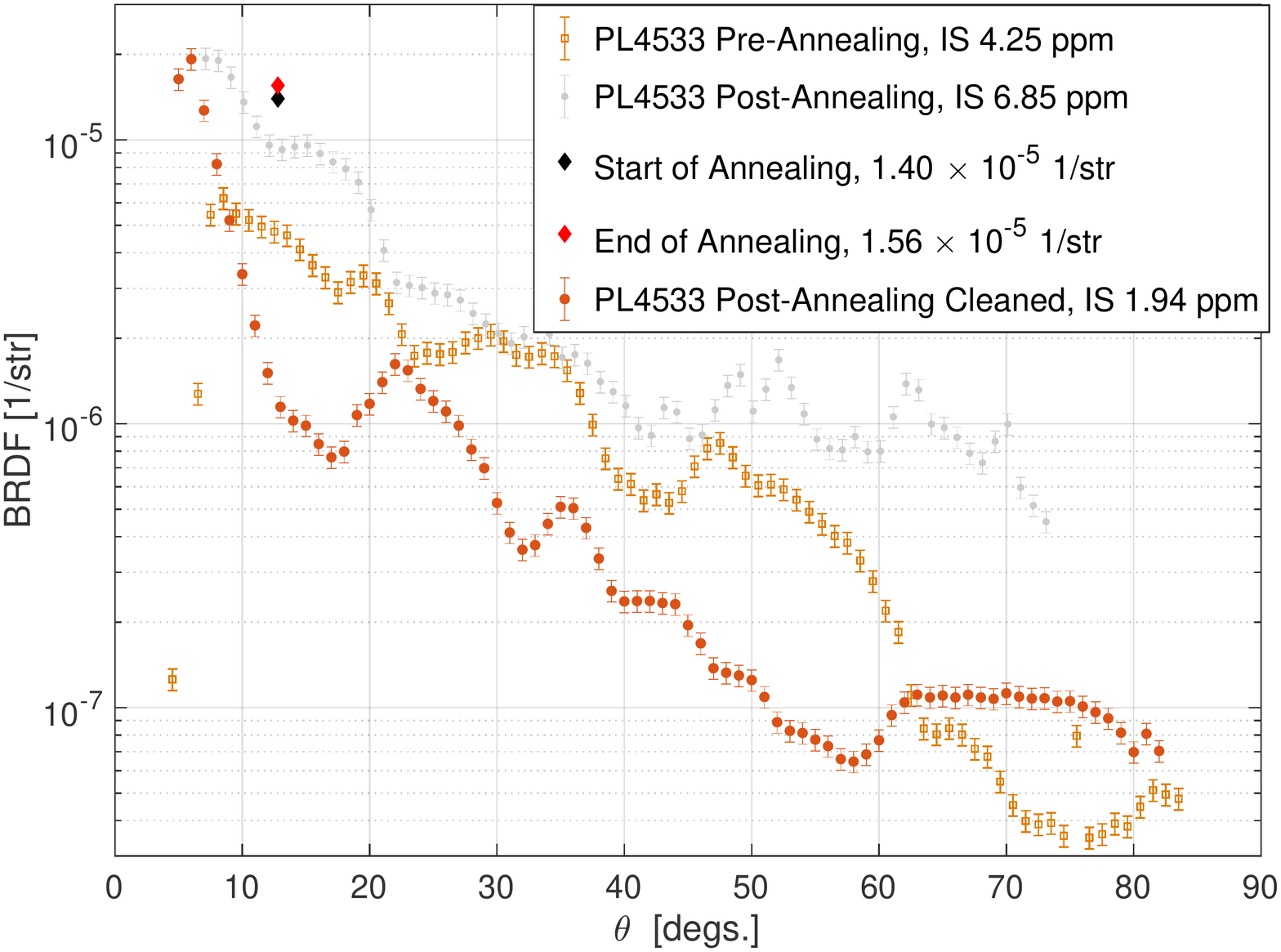}
\end{minipage}
\begin{minipage}{0.5\textwidth}
    \includegraphics[width=1.0\textwidth]{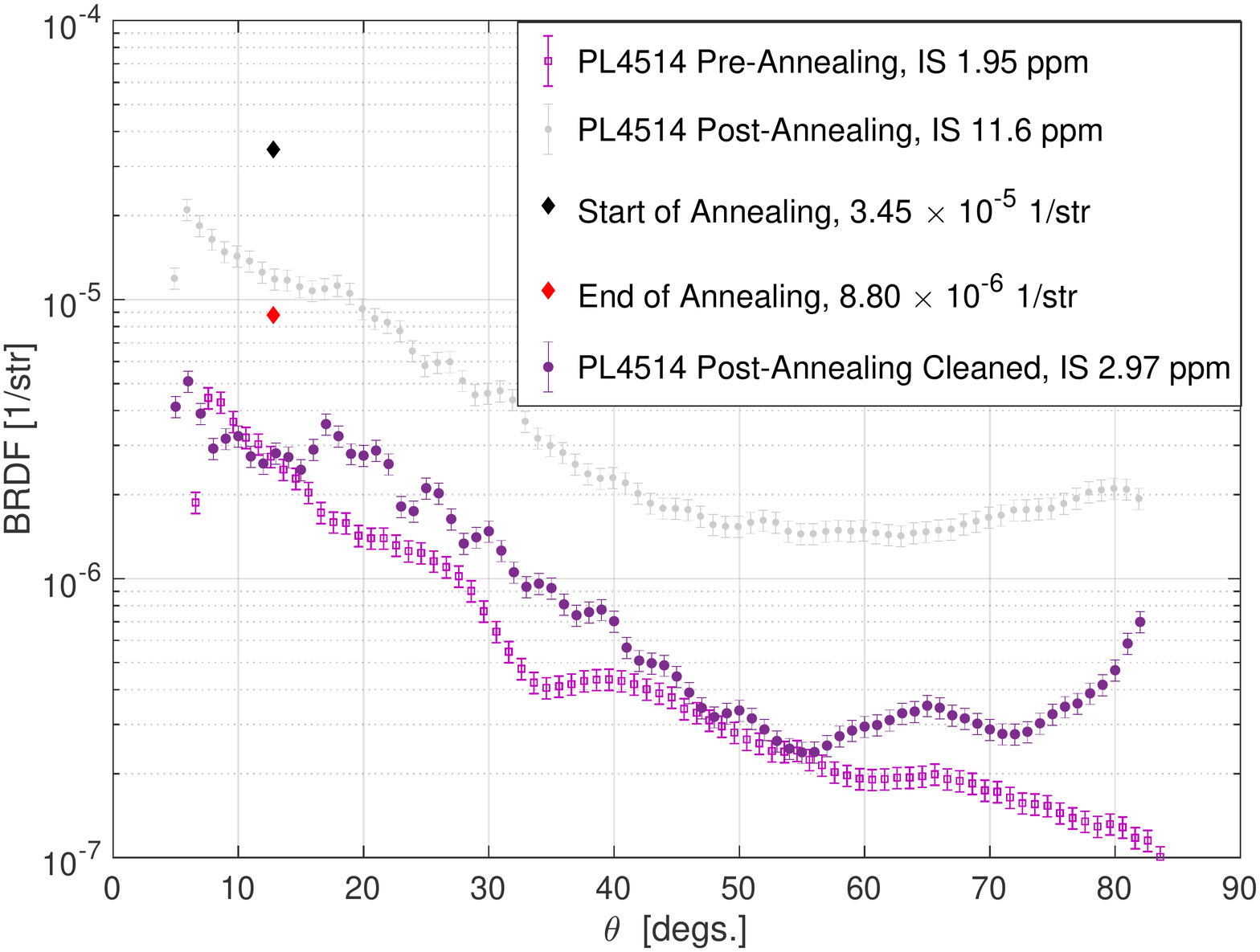}
\end{minipage}
\begin{minipage}[t]{0.5\textwidth}
    \includegraphics[width=1.0\textwidth]{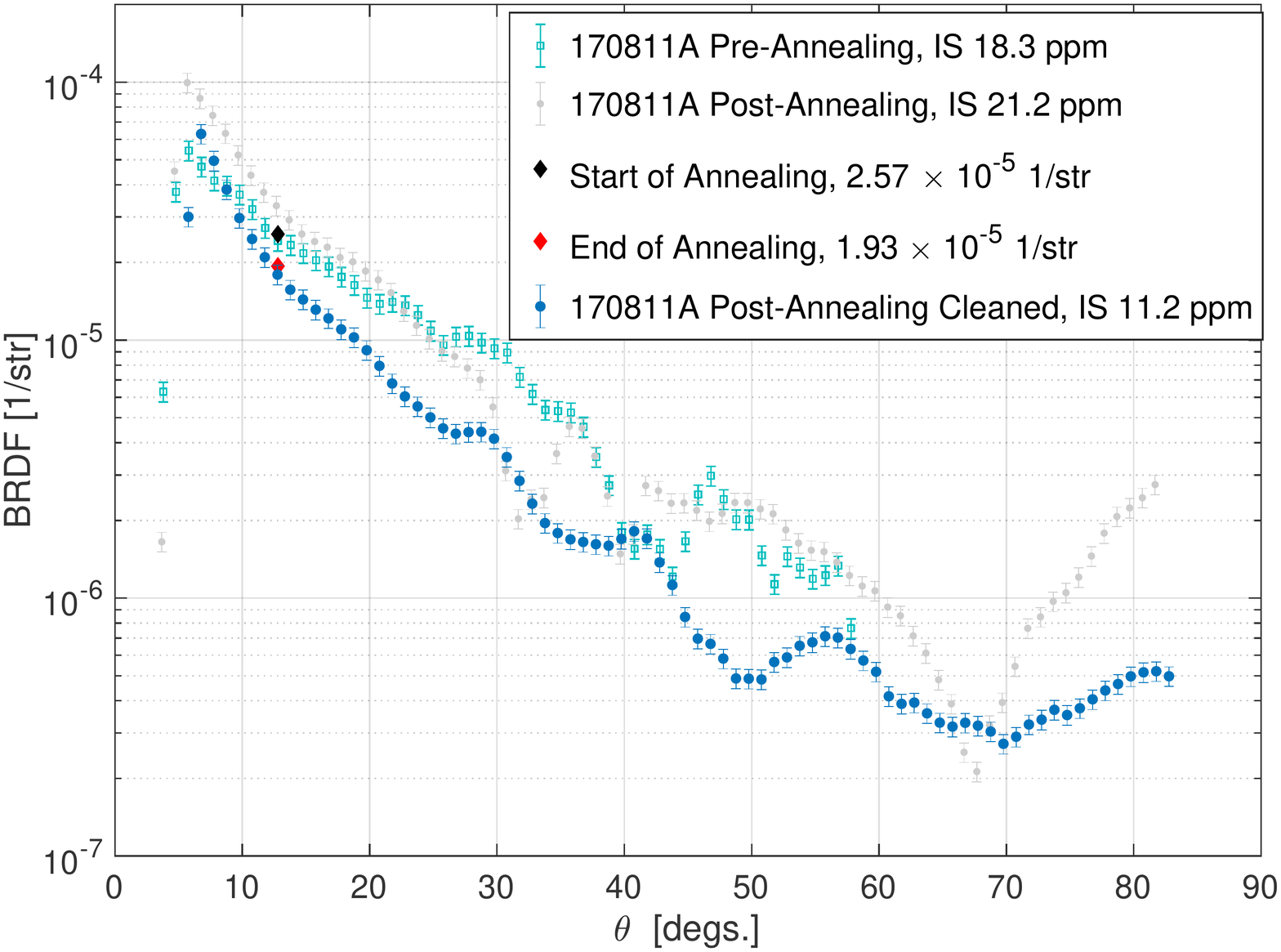}
\end{minipage}
\begin{minipage}[t]{0.5\textwidth}
    \includegraphics[width=1.0\textwidth]{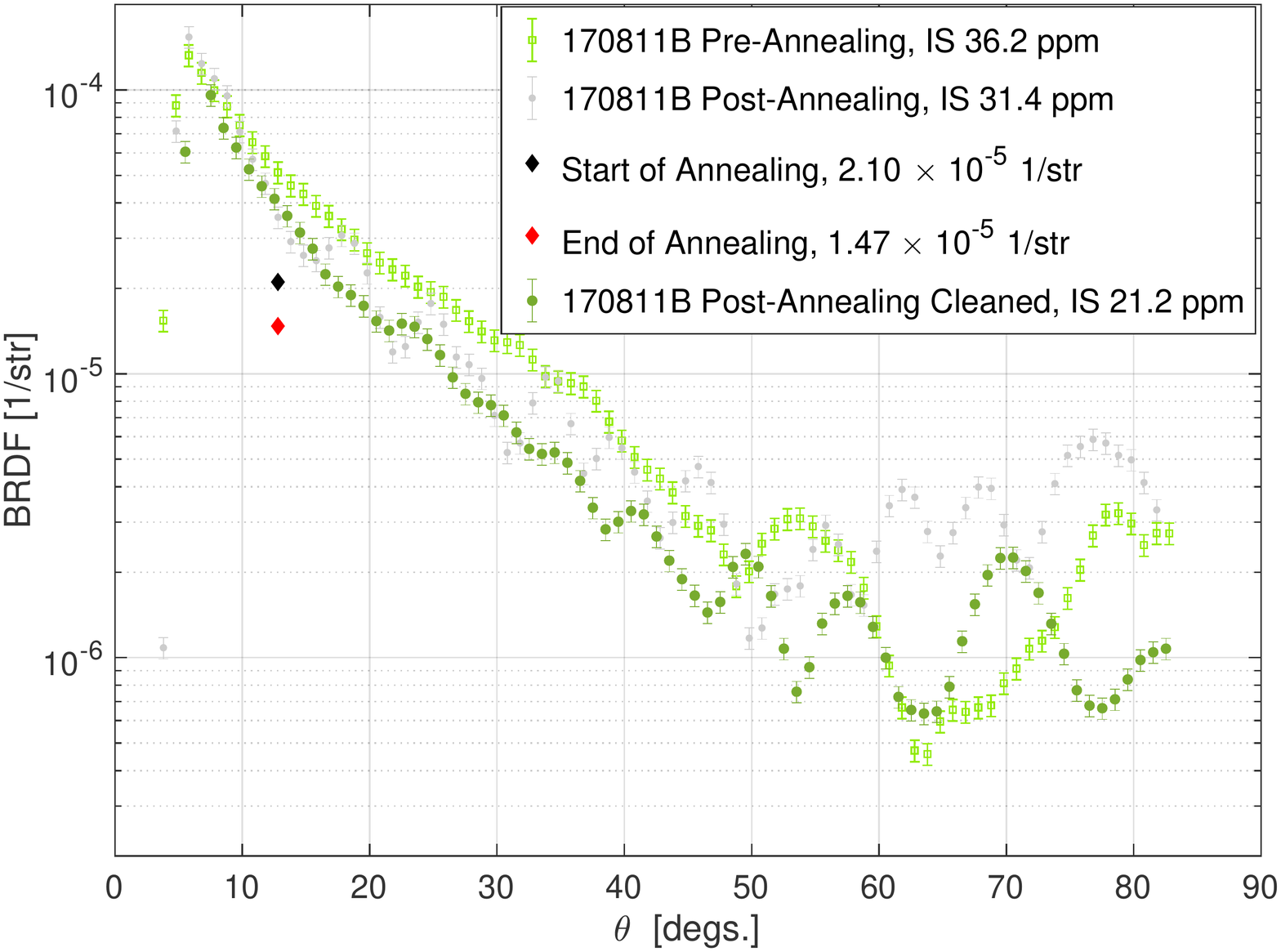}
\end{minipage}
\caption{Scatter versus angle for the tantala-coated optics before and after annealing, measured by ARS. The black and red diamond values represent room temperature BRDF measurements made by the VAS at the start and end of annealing, respectively. The gray curves represent the post-annealing measurement before cleaning. Top left: PL4533. After annealing to 450$^{\circ}$C and cleaning, scatter has decreased over most angles. Top right: PL4514. After annealing to 500$^\circ$C and cleaning, this sample shows only slight increase over all angles. Bottom left: CSU 170811a. After annealing to 490$^\circ$C and cleaning, scatter is somewhat lower for all the angles measured. Due to high scatter from the back of the sample, the pre-annealing measurement is cut off before 58$^\circ$. Bottom right: CSU 170811b. After annealing to 500$^\circ$C and cleaning, scatter is fairly consistent, but somewhat lower, over most angles. Note: pre-annealing results are the same as in Figure~\ref{fig:pre-anneal}.}
\label{fig:pre-post-comparison}
\end{figure}

After VAS measurements the vacuum chamber was opened to (unfiltered room) air and the samples were removed and transported to the ARS experiment. Upon measurement we found that several samples showed anomalously high scatter (see Figure~\ref{fig:pre-post-comparison}) which was incongruous with the decreased scatter seen in the VAS experiment. To check whether this could be from dust or other contamination associated with venting and transport, the front and back surfaces of the optics were cleaned using FirstContact and remeasured in the ARS experiment. The results from these cleaned samples, also presented in  Figure~\ref{fig:pre-post-comparison}, remove some of the anomalously high scatter, and overall improve the scatter of each sample. The need for this process indicates that the samples became dirty sometime after annealing in the VAS and before installation in the ARS setup. 

\begin{center}
    \begin{table}[hb]
        \centering
        \begin{tabular}{c|c|c|c|c}
        \hline
             Sample & T$_{anneal}$ & BRDF 12.8$^{\circ}$ Pre | Post ($10^{-6}$str$^{-1}$) & IS 5-82$^{\circ}$ Pre | Post (ppm) & R Pre | Post \\
             \hline \hline
             PL4533 & 450$^\circ$C & 5.49 | 1.15 & 4.25 | 1.94 &32$\pm$3\% | 20$\pm$2\%   \\
             PL4514 & 500$^\circ$C & 2.74 | 2.80  & 1.95 | 2.97 &28$\pm$3\% | 28$\pm$3\%  \\
             170811a & 490$^\circ$C & 36.9 | 17.9  & 18.3 | 11.2 &13$\pm$1\% | 25$\pm$2\%  \\
             170811b & 500$^\circ$C & 51.2 | 36.0 & 36.2 | 21.2 &6.9$\pm$0.7\% | 12$\pm$1\% \\
        \end{tabular}
        \caption{A summary of the four tantala-coated samples before and after annealing, as measured by ARS. The BRDF is given at a scattering angle of 12.8$^{\circ}$, the angle where VAS measurements are made. The IS is estimated from integrating BRDF over the partial hemisphere of backscattering $5^{\circ} < \theta_s < 82^{\circ}$, assuming isotropic scattering in azimuthal angles~\cite{Magana-Sandoval:12} as described in Section~\ref{sec:analysis}. Reflectivity (R) values were measured using a Thorlabs S121C sensor with $\pm7\%$ accuracy at 1064\,nm. Reflectivity of the uncoated substrate is 13$\pm$1\%.}
        \label{tab:post-sample-info}
    \end{table}
\end{center}

\subsection{LMA Samples}

LMA PL4533 was annealed to 450$^\circ$C. After annealing, the optic was held at room temperature for another day to observe if any change in scatter occurs after the annealing process. The trend in Figure~\ref{fig:annealing_brdf} shows essentially no change in scatter occurred during annealing and while being held at room temperature. After holding the optic at room temperature, it was transported between experiments and again imaged in the angle-resolved scatterometer. Figure~\ref{fig:pre-post-comparison} compares the pre- and post-annealing scatter (before and after cleaning) for this sample. The scatter after annealing and cleaning is lower overall, especially demonstrated by the integrated scatter, which has decreased from 4.25 ppm to 1.94 ppm. Cleaning this sample after annealing was important to preserve the low scatter that was witnessed during VAS measurements.

LMA PL4514 demonstrates the most dramatic BRDF change during its annealing in Figure~\ref{fig:annealing_brdf}, decreasing from 35 ($\pm$3) to 8.8 ($\pm$0.7) $\times 10^{-6}$ str$^{-1}$. During initial alignment of the sample in the VAS, a bright scatter point was observed, and unsuccessful attempts were made to remove it with pressurized nitrogen. This scatter point was not seen in the ARS measurements. The scatter point gradually disappeared during the annealing ramp and was fully gone by 465$^{\circ}C$, which most likely causes the highest contribution to the dramatic scatter decrease. The cause of the bright scatter point is unknown. It is possible that it was water or other organic material that evaporated at elevated temperatures. Pre- and post-annealing ARS data show that the scatter overall has held consistent below 15$^{\circ}$, and increased slightly at higher angles after annealing (IS increased from 1.95 to 2.97 ppm).


\subsection{CSU Samples}

CSU 170811a was annealed to 490$^\circ$C. The start and end values of the scatter, measured at room temperature with the VAS, decreased from 2.6 ($\pm$0.2) to 1.9 ($\pm$0.2) $\times$ 10$^{-5}$ str$^{-1}$ (Figure~\ref{fig:annealing_brdf}). Comparing the pre- and post-annealing measurements (Figure~\ref{fig:pre-post-comparison}) shows consistent scatter at low angles. A determination of the high angle scatter in the pre-annealing measurement could not be made due to scattering from the back of the sample contaminating the region of interest. Scatter appears to have decreased for this sample overall; the final IS is 11.2 ppm, compared to an initial value of 18.3 ppm up to 58$^{\circ}$. Because the pre-annealing measurement truncates at this lower angle, it cannot be determined if there is a change in scatter at high angles in the post-annealing measurement.

CSU 170811b's start and end room temperature values, measured in VAS, show some decrease: from 2.1 ($\pm$0.2) to 1.4 ($\pm$0.1) $\times$10$^{-5}$ str$^{-1}$ (Figure~\ref{fig:annealing_brdf}). Annealing to 500$^\circ$C reduced the integrated scatter from 36.2 ppm to 21.2 ppm. Furthermore, the scatter at 12.8$^\circ$ measured in ARS decreased, from 5.12 to 3.60 $\times$10$^{-5}$ str$^{-1}$ (as seen in Tables~\ref{tab:sample-info} and \ref{tab:post-sample-info}). 
The post-annealing scatter also shows an interesting trend of bumps versus scattering angle. This contrasts the smooth trend in the pre-annealing scatter measurement. Such periodic behavior can come from point scatterers~\cite{Lequime2008}.

\section{Conclusions and Future Work}

These results demonstrate that post-deposition high-temperature annealing of single layer tantala and titania-doped tantala thin films in vacuum does not lead to an increase in scatter, and may actually improve their scatter. 
Images of the samples during the annealing process show that scatter points disappear, suggesting that the improvement could be caused by evaporation of water or other material on the face of the optics or by coating material changes. 
Given the apparent repeatability of this phenomenon, it is interesting to consider whether in-vacuum post-deposition annealing of the LIGO-Virgo core optics could decrease their scatter. As shown in Figure~\ref{fig:annealing_brdf}, it may not be necessary to reach 500$^{\circ}$C to achieve the benefits. However, it is also possible that any scattering reduction has already been achieved by the in-air annealing of the LIGO-Virgo optics.
Although these results look promising for reducing scatter, no measurements have been made to indicate whether other properties such as absorption loss have changed. Additionally, these studies focused on single layers of the high-index material and may not be representative of the full coating stacks used by LIGO-Virgo.
Since this experiment has shown promising results in vacuum, the next step is to develop a similar apparatus that monitors scatter with in-air annealing.


Some samples, in initial post-annealing measurements, showed anomalous high scatter. Cleaning these samples decreased this anomalous scatter, especially for the LMA samples. The CSU samples did not have the same large scatter differences that the LMA samples did, but cleaning them still improved the post-annealing scatter. Most samples showed some decrease in scatter during annealing at vacuum, so it appears that the process of bringing samples to atmospheric pressure and transporting them between experiments could introduce dirt or other impurities. Therefore, cleaning with FirstContact was necessary to maintain the decrease in scatter witnessed during annealing. Investigations are underway to find and eliminate the source of the contamination. 

Figure~\ref{fig:pre-post-comparison} also showed a discrepancy between the BRDF values measured in the pre-annealing ARS and the start of the VAS experiments, which should nominally have been similar. These differences could be due to the different beam sizes and spot placement in each apparatus, which were adjusted by hand. It could also be caused by the different coherent scattering effects caused by the different coherence lengths of the laser used in ARS and the SLED used in VAS. This is an area for future investigation. 


The temperatures used in this study were not high enough to cause crystallization changes in the samples. Future work involves increasing the annealing temperature closer to the crystallization point of tantala (650$^{\circ}$C). Also, while this study focused on titania-doped tantala due to its use in aLIGO mirrors, other thin films under research for use in aLIGO+ such as germania (GeO$_2$), or coatings with different applications, could be studied in the vacuum annealing scatterometer.

\section{Acknowledgements}
The authors benefited greatly from conversations with members of the LIGO Scientific Collaboration and especially the LIGO Center for Coatings Research. We thank Ashot Markosyan for guidance on building the vacuum annealing scatterometer and Rana Adhikari for useful discussions regarding the experiment. We thank Mariana Fazio and Riccardo DeSalvo for careful review of the manuscript. This work was supported by NSF grants PHY-1708035, PHY-1807069, and AST-1559694. EC was supported in part by the Nancy Goodhue-McWilliams Graduate Fellowship. JS and AG were supported in part by the Dan Black Family Trust. Amorphous thin films of Ta$_2$O$_5$ were deposited at Colorado State University by Le Yang and Mariana Fazio in the group of Carmen Menoni with the support of the NSF/Moore Foundation Center for Coatings Research.
\bibliography{Scatter}

\begin{thebibliography}{10}
\newcommand{\enquote}[1]{``#1''}

\bibitem{Harry:2011book}
G.~Harry, T.~P. Bodiya, and R.~DeSalvo, eds., \emph{Optical Coatings and
  Thermal Noise in Precision Measurement} (Cambridge University Press, 2011).

\bibitem{gw150914}
B.~P. Abbott \emph{et~al.}, \enquote{Observation of gravitational waves from a
  binary black hole merger,} {\protect\JournalTitle{Phys. Rev. Lett.}}
  \textbf{116}, 061102 (2016).

\bibitem{gw170817}
B.~P. Abbott \emph{et~al.}, \enquote{Gw170817: Observation of gravitational
  waves from a binary neutron star inspiral,} {\protect\JournalTitle{Phys. Rev.
  Lett.}} \textbf{119}, 161101 (2017).

\bibitem{LIGOScientific:2018mvr}
B.~P. Abbott \emph{et~al.}, \enquote{{GWTC-1: A Gravitational-Wave Transient
  Catalog of Compact Binary Mergers Observed by LIGO and Virgo during the First
  and Second Observing Runs},} {\protect\JournalTitle{Phys. Rev.}} \textbf{X9},
  031040 (2019).

\bibitem{TheLIGOScientific:2014jea}
J.~Aasi \emph{et~al.}, \enquote{{Advanced LIGO},} {\protect\JournalTitle{Class.
  Quant. Grav.}} \textbf{32}, 074001 (2015).

\bibitem{TheVirgo:2014hva}
F.~Acernese \emph{et~al.}, \enquote{{Advanced Virgo: a second-generation
  interferometric gravitational wave detector},} {\protect\JournalTitle{Class.
  Quant. Grav.}} \textbf{32}, 024001 (2015).

\bibitem{coreoptics}
G.~Billingsley, H.~Yamamoto, and L.~Zhang, \enquote{Characterization of
  advanced {LIGO} core optics,} {\protect\JournalTitle{Proc. of ASPE}}
  \textbf{66}, 78--83 (2017).

\bibitem{PhysRevD.98.122001}
S.~Gras and M.~Evans, \enquote{Direct measurement of coating thermal noise in
  optical resonators,} {\protect\JournalTitle{Phys. Rev. D}} \textbf{98},
  122001 (2018).

\bibitem{Granata_2020}
M.~Granata, A.~Amato, L.~Balzarini, M.~Canepa, J.~Degallaix, D.~Forest,
  V.~Dolique, L.~Mereni, C.~Michel, L.~Pinard, B.~Sassolas, J.~Teillon, and
  G.~Cagnoli, \enquote{Amorphous optical coatings of present gravitational-wave
  interferometers,} {\protect\JournalTitle{Classical and Quantum Gravity}}
  \textbf{37}, 095004 (2020).

\bibitem{Amato_2018}
A.~Amato, G.~Cagnoli, M.~Canepa, E.~Coillet, J.~Degallaix, V.~Dolique,
  D.~Forest, M.~Granata, V.~Martinez, C.~Michel, L.~Pinard, B.~Sassolas, and
  J.~Teillon, \enquote{High-reflection coatings for gravitational-wave
  detectors:state of the art and future developments,}
  {\protect\JournalTitle{Journal of Physics: Conference Series}} \textbf{957},
  012006 (2018).

\bibitem{Fazio:2020yqa}
M.~A. Fazio, G.~Vajente, A.~Ananyeva, A.~Markosyan, R.~Bassiri, M.~M. Fejer,
  and C.~S. Menoni, \enquote{Structure and morphology of low mechanical loss
  tio2-doped ta2o5,} {\protect\JournalTitle{Opt. Mater. Express}} \textbf{10},
  1687--1703 (2020).

\bibitem{Accadia_2010}
T.~Accadia, F.~Acernese, F.~Antonucci, P.~Astone, G.~Ballardin, F.~Barone,
  M.~Barsuglia, T.~S. Bauer, M.~G. Beker, A.~Belletoile, S.~Birindelli,
  M.~Bitossi, M.~A. Bizouard, M.~Blom, F.~Bondu, L.~Bonelli, R.~Bonnand,
  V.~Boschi, L.~Bosi, B.~Bouhou, S.~Braccini, C.~Bradaschia, A.~Brillet,
  V.~Brisson, R.~Budzy{\'{n}}ski, T.~Bulik, H.~J. Bulten, D.~Buskulic, C.~Buy,
  G.~Cagnoli, E.~Calloni, E.~Campagna, B.~Canuel, F.~Carbognani, F.~Cavalier,
  R.~Cavalieri, G.~Cella, E.~Cesarini, E.~C. Mottin, A.~Chincarini, F.~Cleva,
  E.~Coccia, C.~N. Colacino, J.~Colas, A.~Colla, M.~Colombini, A.~Corsi, J.-P.
  Coulon, E.~Cuoco, S.~D{\textquotesingle}Antonio, V.~Dattilo, M.~Davier,
  R.~Day, R.~D. Rosa, G.~Debreczeni, M.~del Prete, L.~D. Fiore, A.~D. Lieto,
  M.~D.~P. Emilio, A.~D. Virgilio, A.~Dietz, M.~Drago, V.~Fafone, I.~Ferrante,
  F.~Fidecaro, I.~Fiori, R.~Flaminio, J.-D. Fournier, J.~Franc, S.~Frasca,
  F.~Frasconi, A.~Freise, M.~Galimberti, L.~Gammaitoni, F.~Garufi, M.~E.
  G{\'{a}}sp{\'{a}}r, G.~Gemme, E.~Genin, A.~Gennai, A.~Giazotto, R.~Gouaty,
  M.~Granata, C.~Greverie, G.~M. Guidi, J.-F. Hayau, H.~Heitmann, P.~Hello,
  S.~Hild, D.~Huet, P.~Jaranowski, I.~Kowalska, A.~Kr{\'{o}}lak, N.~Leroy,
  N.~Letendre, T.~G.~F. Li, M.~Lorenzini, V.~Loriette, G.~Losurdo, E.~Majorana,
  I.~Maksimovic, N.~Man, M.~Mantovani, F.~Marchesoni, F.~Marion, J.~Marque,
  F.~Martelli, A.~Masserot, C.~Michel, L.~Milano, Y.~Minenkov, M.~Mohan,
  N.~Morgado, A.~Morgia, S.~Mosca, V.~Moscatelli, B.~Mours, I.~Neri, F.~Nocera,
  G.~Pagliaroli, L.~Palladino, C.~Palomba, F.~Paoletti, S.~Pardi, M.~Parisi,
  A.~Pasqualetti, R.~Passaquieti, D.~Passuello, G.~Persichetti, M.~Pichot,
  F.~Piergiovanni, M.~Pietka, L.~Pinard, R.~Poggiani, M.~Prato, G.~A. Prodi,
  M.~Punturo, P.~Puppo, D.~S. Rabeling, I.~R{\'{a}}cz, P.~Rapagnani, V.~Re,
  T.~Regimbau, F.~Ricci, F.~Robinet, A.~Rocchi, L.~Rolland, R.~Romano,
  D.~Rosi{\'{n}}ska, P.~Ruggi, B.~Sassolas, D.~Sentenac, L.~Sperandio,
  R.~Sturani, B.~L. Swinkels, A.~Toncelli, M.~Tonelli, O.~Torre, E.~Tournefier,
  F.~Travasso, G.~Vajente, J.~F.~J. van~den Brand, S.~van~der Putten,
  M.~Vasuth, M.~Vavoulidis, G.~Vedovato, D.~Verkindt, F.~Vetrano,
  A.~Vicer{\'{e}}, J.-Y. Vinet, H.~Vocca, M.~Was, and M.~Yvert, \enquote{Noise
  from scattered light in virgo{\textquotesingle}s second science run data,}
  {\protect\JournalTitle{Classical and Quantum Gravity}} \textbf{27}, 194011
  (2010).

\bibitem{Flanagan_1994}
E.~Flanagan and K.~S. Thorne, \enquote{Noise due to backscatter off baffles,
  the nearby wall, and objects at the far end of the beam tube; and recommended
  actions,} Tech. Rep. LIGO-T940063-00-R, California Institute of Technology
  (1994).

\bibitem{Kwee:2014vba}
P.~Kwee, J.~Miller, T.~Isogai, L.~Barsotti, and M.~Evans, \enquote{{Decoherence
  and degradation of squeezed states in quantum filter cavities},}
  {\protect\JournalTitle{Phys. Rev. D}} \textbf{90}, 062006 (2014).

\bibitem{Smith:19}
J.~R. Smith, R.~X. Adhikari, K.~M. Aleman, A.~Avila-Alvarez, G.~Billingsley,
  A.~Gleckl, J.~Guerrero, A.~Markosyan, S.~Penn, J.~A. Rocha, D.~Rose, and
  R.~Wright, \enquote{Apparatus to measure optical scatter of coatings versus
  annealing temperature,} in \emph{Optical Interference Coatings Conference
  (OIC) 2019,}  (Optical Society of America, 2019), p. FA.2.

\bibitem{Glover:2018huy}
L.~Glover \emph{et~al.}, \enquote{{Optical scattering measurements and
  implications on thermal noise in Gravitational Wave detectors test-mass
  coatings},} {\protect\JournalTitle{Phys. Lett. A}} \textbf{382}, 2259--2264
  (2018).

\bibitem{Glover_2018}
L.~Glover, M.~Goff, S.~Linker, J.~Neilson, J.~Patel, I.~Pinto, M.~Principe,
  E.~Villarama, E.~Arriaga, E.~Barragan, S.~Chao, L.~Daneshgaran, R.~DeSalvo,
  E.~Do, and C.~Fajardo, \enquote{A multi-step approach to assessing {LIGO}
  test mass coatings,} {\protect\JournalTitle{Journal of Physics: Conference
  Series}} \textbf{957}, 012010 (2018).

\bibitem{Stover:2012book}
J.~C. Stover, \emph{Optical Scattering} (SPIE Press, 2012), 3rd ed.

\bibitem{Lequime2008}
M.~Lequime, M.~Zerrad, C.~Deumi\'e, and C.~Amra, \enquote{A goniometric light
  scattering instrument with high-resolution imaging,}
  {\protect\JournalTitle{Optics Communications}}  (2008).

\bibitem{schroder2011angle}
S.~Schr{\"o}der, T.~Herffurth, H.~Blaschke, and A.~Duparr{\'e},
  \enquote{Angle-resolved scattering: an effective method for characterizing
  thin-film coatings,} {\protect\JournalTitle{Applied optics}} \textbf{50},
  C164--C171 (2011).

\bibitem{Amra:2004cqa}
F.~Beauville \emph{et~al.}, \enquote{{Low-loss coatings for the VIRGO large
  mirrors},} {\protect\JournalTitle{Proc. SPIE Int. Soc. Opt. Eng.}}
  \textbf{5250}, 483--492 (2004).

\bibitem{Magana-Sandoval:12}
F.~Magana-Sandoval, R.~X. Adhikari, V.~Frolov, J.~Harms, J.~Lee, S.~Sankar,
  P.~R. Saulson, and J.~R. Smith, \enquote{Large-angle scattered light
  measurements for quantum-noise filter cavity design studies,}
  {\protect\JournalTitle{J. Opt. Soc. Am. A}} \textbf{29}, 1722--1727 (2012).

\bibitem{vander-hyde2015}
D.~Vander-Hyde, C.~Amra, M.~Lequime, F.~Magaña-Sandoval, J.~R. Smith, and
  M.~Zerrad, \enquote{Optical scatter of quantum noise filter cavity optics,}
  {\protect\JournalTitle{Classical and Quantum Gravity}} \textbf{32}, 135019
  (2015).

\end{thebibliography}

\end{document}